Report on the ESO Workshop

# Inward Bound: Bulges from High Redshifts to the Milky Way

held online, 2–6 May 2022

Dimitri Gadotti[1]
Elena Valenti[2]
Francesca Fragkoudi[3]
Anita Zanella[4]
Lodovico Coccato[2]
Camila de Sá Freitas[2]
Stella-Maria Chasiotis-Klingner[2]

[1] Centre for Extragalactic Astronomy, Department of Physics, Durham University
[2] ESO
[3] Institute for Computational Cosmology, Department of Physics, Durham University
[4] National Institute of Astrophysics, Padova

With over 200 registered participants, this fully online conference allowed theorists and observers across the globe to discuss recent findings on the central structures of disc galaxies. By design, this conference included experts on the Milky Way, local and high-redshift galaxies, and theoretical aspects of galaxy formation and evolution. The need for such a broad range of expertise stems from the important advances that have been made on all fronts in recent years. One of the main goals of this meeting was accordingly to bring together these different communities, to find a common ground for discussion and mutual understanding, to exchange ideas, and to efficiently communicate progress.

Like many other meetings, this conference had to be postponed twice since 2020 because of the COVID-19 pandemic. Although the original plan was to have an in-person meeting, both the Scientific Organising Committee (SOC) and the Local Organising Committee (LOC) felt that further postponing the conference would be too detrimental. During the two years of the pandemic, the LOC had gained sufficient experience to be able to devise a format that would facilitate the meeting's intended goal of fostering discussions between the different communities.

The meeting consisted of 31 pre-recorded talks (made available to registered participants a week before the start of the conference), as well as six live sessions, held on the Monday, Wednesday, and Friday, which included 12 invited talks, four review talks, and four discussion sessions[1]. The live sessions took place in the morning and early evening in Europe, to enable the participation of colleagues from time zones in the Americas and Australia/Asia. Those sessions were recorded and made available immediately afterwards. This allowed participants in different time zones to be up to date with all the live sessions, while a Slack workspace allowed the participants to have further asynchronous interactions and discussions. We were pleased to see that this setup worked very well in fostering numerous and deep discussions as intended. The pre-recorded talks and recordings of the live sessions are now publicly available to the community[2]. In what follows we summarise some of the main discussion topics and outcomes of the workshop.

## Nomenclature

An overarching discussion concerned the topic of nomenclature. We confirmed that the term 'bulge' was used to indicate different physical structures by different research teams and communities. It was generally accepted that this situation is detrimental and that raising awareness of this issue is a first step towards a solution. It was thus generally agreed that, to clearly communicate results, it is important to define from the onset what the employed nomenclature means in terms of physical structures and associated properties.

In statistical studies using large samples, the term 'bulge' usually indicates any structure within the inner kiloparsec which is not the main disc. Similarly, the term 'photometric bulge', typically employed in the context of photometric decompositions, may encompass more than one physical structure (for example, a bar and a nuclear disc). As discussed in talks such as those by Simon Driver, Adriana de Lorenzo-Cáceres and Marie Martig, the main central stellar structures often found in disc galaxies are: (i) a pressure-supported (yet, rotating) spheroid; (ii) the inner part of a bar that grows out of the disc plane and shows a boxy or peanut/X-like morphology; and (iii) a rotation-supported disc (which is not the main, large-scale galaxy disc). Respectively, they are often referred to as the 'classical bulge' (CB), 'boxy/peanut bulge' (BP) and 'nuclear disc' (ND). Some participants argued for dropping the word 'bulge' from the second and replacing 'nuclear' by 'inner' in the third. The term 'pseudo-bulge', which can refer to both BPs and NDs was said by many participants to be particularly confusing, despite being widely used.

## The Milky Way

A consensus has been reached that the Milky Way has primarily a BP formed from the bar, with stellar populations born in situ (see talks by Paola Di Matteo, Francesca Fragkoudi and Melissa Ness). However, there is still a healthy debate around the ages of the stellar populations in the central regions (see talks by Tommaso Marchetti, Michael Rich, Álvaro Rojas-Arriagada and Manuela Zoccali). The contribution of the halo to the old population is clear but it is still difficult to quantify. This leaves an open question: is there still space for a low-mass, old, central spheroidal structure that is not part of the halo? In other words, is there room yet for a CB in the Milky Way? (See the talks by Cristina Chiappini and Madeline Lucey.)

In this context, we still lack a comprehensive characterisation of the most metal-poor population (with [Fe/H] < –1), from both the modelling and observational sides, even though significant progress has recently been made (see talks by Anke Arentsen, Andrea Kunder, Giulia Pagnini and Jason Sanders).

## Formation scenarios

While it is still unclear what is the physical mechanism that produces BPs from the inner parts of bars (i.e., whether they form from buckling instabilities or orbital resonances), it is well established that BPs are simply the vertically thicker inner parts of bars (see talks by Sandor Kruk and Jairo Méndez-Abreu).

Likewise, there is mounting evidence that NDs form via gas inflow produced by bar-driven processes (see talks by Dimitri

Gadotti, Camila de Sá Freitas, Patricia Sánchez-Blázquez and Mattia Sormani). Nevertheless, the possibility of forming NDs via processes driven by clumps in discs at high redshift was also discussed.

A topic that sparked great interest was the possibility of forming CBs through clump-driven processes. Numerical simulations show clearly that if clumps survive feedback processes they coalesce to build a central pressure-supported spheroid; observations provide some support for this scenario (see talks by Daniel Ceverino, Miroslava Dessauges-Zavadsky, Deanne Fisher, Yicheng Guo, Thorsten Naab, Stijn Wuyts and Anita Zanella). The survival of the largest clumps can also provide constraints on how feedback works. On the other hand, while some participants argued that the clump scenario is favoured over the merger scenario to form most CBs, arguments concerning the sizes and masses of clumps suggest that mergers may still be important in forming the most massive CBs. Nevertheless, a challenge to the clump scenario is brought about by the difficulty of detecting clumps in the molecular gas distribution. Paradoxically, this would indicate high star formation efficiencies, leading to strong feedback, which in turn would dissolve clumps before they can migrate to the centre. Further, separating the effects of mergers and clump migration is still difficult (see talks by Marc Huertas-Company and Annagrazia Puglisi) and whether clumps can contribute to the formation of thick discs remains an open question. A general consensus was that more theoretical work is needed to understand the evolution of clumps and their contribution to bulge formation (for example, what is the expected range of physical properties such as the sizes and masses of bulges formed from the coalescence of clumps?).

These, and alternative, formation scenarios were also discussed in talks by Francesco Ferraro, Hua Gao, Yicheng Guo, Wako Ishibashi, Keerthana Jegatheesan, Kalina Nedkova, Sandro Tacchella and Jesse van de Sande). To some extent, a typical disc galaxy may go through all of these different processes (i.e., clump migration, mergers, bar-driven processes etc.), leading to the formation of composite bulges, where CBs, BPs and NDs co-exist (see the talk by Peter Erwin). In this context, participants also discussed that it is not clear if there is a line to be drawn between CBs and NDs. In fact, in some cases it is difficult to link observed physical properties to a single formation scenario, which can be an indication that several mechanisms play important roles in the formation of these central stellar structures.

In talks such as those by Ignacio Gargiulo, Aura Obreja and Milena Valentini, it was shown that simulations are producing realistic disc-dominated systems, although it is still not clear if they quantitatively reproduce observations. Comparing measurements from observations and simulations requires not only homogenous metrics but also understanding the systematics on both fronts, which is not trivial. In this context, a remark made by several participants concerned the apparent scarcity of massive CBs in nearby galaxies. This is in contrast with statistical studies of galaxies at intermediate redshifts, which find that CBs dominate in mass (but not in number), although it was noted that such studies suffer from relatively poorer physical spatial resolution.

A crucial piece of information comes from dynamical studies at high redshift, which find CBs (or proto-CBs) at $z \sim 1$–$5$, hosted by kinematically cold discs. This argues that at least some bulges and discs are already in place early on (see the talks by Luca Costantin, Federico Lelli, and Francesca Rizzo).

## Future perspectives

A frequent discussion topic was the power of combining photometric analyses with results from spectroscopic studies, in particular to ascertain the kinematical properties of the stellar populations under investigation.

With a plethora of new facilities coming online in the next 2–10 years or so, the future does look bright for studies of the central regions of disc galaxies. Multi-object spectroscopy surveys to be carried out with, for example, the 4-metre Multi-Object Spectrograph Telescope (4MOST), the Multi-Object Optical and Near-infrared Spectrograph (MOONS) and by the Vera C. Rubin Observatory may in particular shed light on the metal-poor populations in the central region of the Milky Way, which will help address what is the halo contribution therein. Likewise, the expected giant leap in statistical power is bound to offer more thorough views on the nuclear disc of the Galaxy (see the talk by Oscar González).

The High Angular Resolution Monolithic Optical and Near-infrared Integral field spectrograph (HARMONI) at ESO's Extremely Large Telescope will bring with it the possibility of obtaining spatially resolved spectroscopy of the central regions of galaxies at $z > 1$, allowing us to directly study the formation of CBs, BPs and NDs. We will witness the unfolding of the relevant physical processes, which will provide more robust answers to the problems discussed in this conference. Of course, new questions will arise, but overall our understanding will improve, including from serendipitous discoveries that we cannot now even foresee.

## Demographics

The SOC aimed at having a fair representation from the community in terms of scientific interests (i.e., Milky Way, local galaxies, high-redshift galaxies, and theory), gender, geographical regions and to some extent seniority as well. Therefore, for each of the four fields, a reviewer and three invited speakers were proposed based on discussions within the whole SOC. This led to a total of 16 review and invited talks with a 44:56 ratio of female to male speakers. On the other hand, the selection of the 31 contributed talks was performed anonymously, by hiding the author's name and any identifying information. In so doing, as far as the gender balance is concerned, we obtained a good match between abstract submission (49% female, 49% male, 2% non-binary) and talk allocation (52% female, 45% male, 3% non-binary). A fairly good balance between abstract submission and talk allocation was also achieved in terms of geographical regions, with ratios of 56:61 (percentage of submitted and allocated talks) for Europe, 16:19 for US, 17:10 for Central and South America, 3:3 for Asia, and 8:6 for Australia. The adopted anonymous method was fairly successful in producing a balanced programme in the context of gender, geographical location and seniority, with little need for the SOC to adjust the outcome of the votes.

The workshop had a high level of participation with about 230 registered participants attending from all continents but Antarctica, with the following percentages:

– 41% Europe (Belgium, Croatia, Denmark, France, Germany, Greece, Italy, Malta, Poland, Portugal, Spain, Sweden, Switzerland, UK)
– 8% North America (Canada, US)
– 2% Central America (Guatemala, Mexico)
– 8% South America (Argentina, Brazil, Chile)
– 4% Oceania (Australia)
– 36% Asia (China, India, Iran, Japan, South Korea)
– 1% Africa (Etiopia, Nigeria)

By enabling easy and cheap access (there was no registration fee for the conference), the online format was a powerful way to reach a very diverse audience, not only in terms of geographical regions but also of career levels. In fact, about 50% of the participants were students, about 16% were postdoctoral researchers, and about 26% on a tenure track or faculty position (a fraction of participants did not specify their career level).

Acknowledgements

We wholeheartedly thank Tutku Kolcu, Alonso Luna and Nelma Silva for their invaluable support and dedication to making this a successful meeting. We are grateful to the SOC members for their sustained commitment during these difficult times, as well as for their key contribution and insight as chairs of several discussion sessions.

Links

[1] Link to workshop programme: https://www.eso.org/sci/meetings/2022/BULGES2022/program.html
[2] Link to access all workshop recordings (i.e., review, invited and contributed talks, as well as the discussion sessions): https://www.eso.org/sci/meetings/2022/BULGES2022/restricted.html

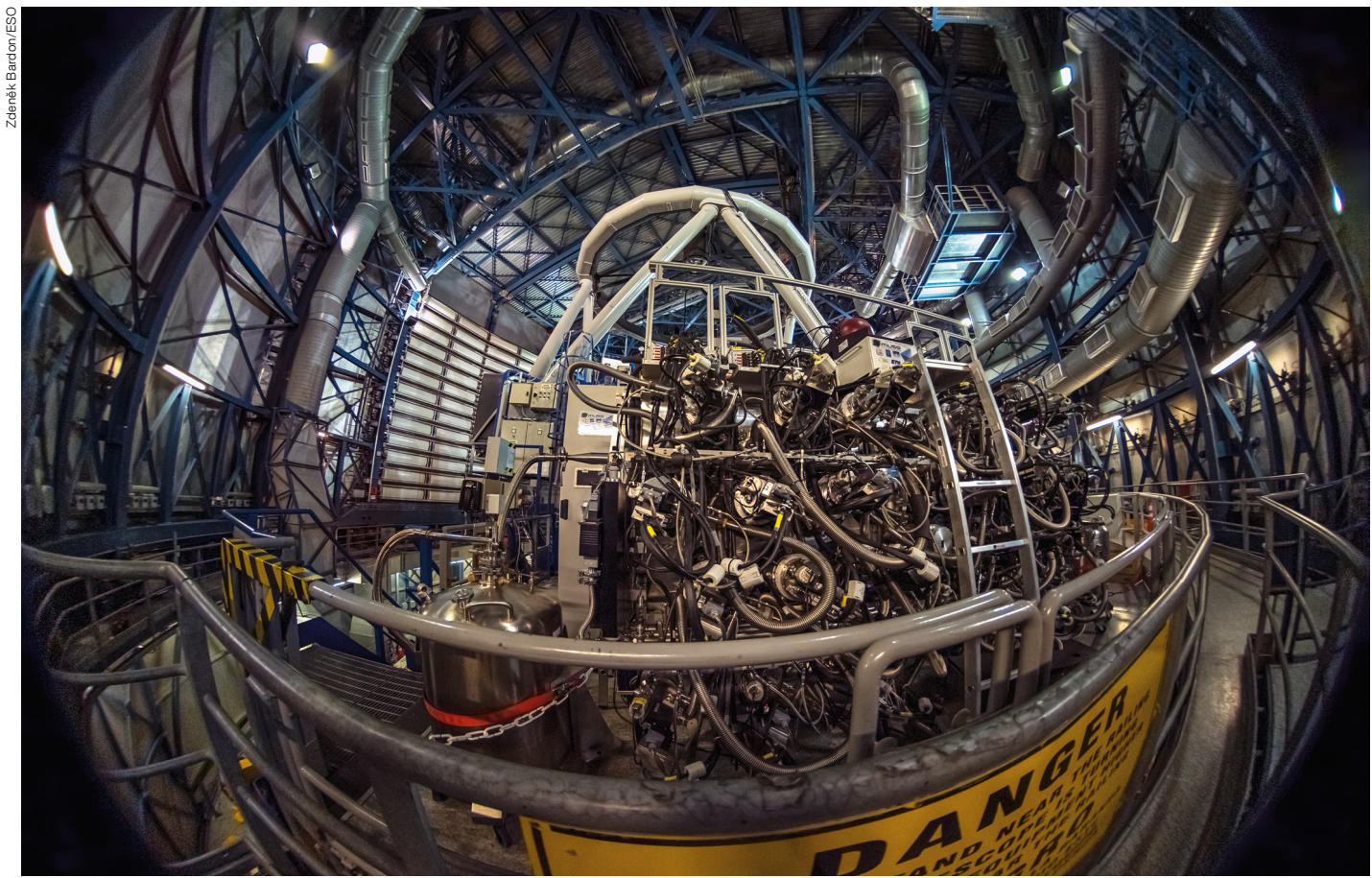

Is this tangle of cords and hoses a machine from the movie The Matrix? You can stay calm: even though the sign says "danger", what may look like a threatening machine is actually the Multi Unit Spectroscopic Explorer (MUSE) instrument on ESO's Very Large Telescope (VLT) at Paranal Observatory. MUSE is one of the largest instruments at the VLT and is connected to one of its four 8.2 m telescopes, Yepun.